\documentclass[12pt,a4paper]{article}
\usepackage{amssymb,amsmath,graphicx,subcaption,hyperref,cite}
\usepackage{epstopdf}
\usepackage[utf8]{inputenc}
\usepackage[english]{babel}
\begin{document}
\begin{center}
   {\Large\bf Transient behaviour towards the stable limit cycle in the Sel'kov model of Glycolysis: A physiological
   disorder}
\end{center}
\vskip 1 cm
\begin{center} 
    Tanmay Das$^{1,*}$ and Muktish Acharyya$^{2,\dagger}$
   
   \textit{${^1}$Department of Physics,Government General Degree College at Kalna-I}\\
   \textit{Muragacha, Medgachi, Purba Bardhdaman, Pin 713405,West Bengal, India}

   \textit{${^2}$Department of Physics, Presidency University,}\\
   \textit{86/1 College Street, Kolkata-700073, India} 
\vskip 0.2 cm
   
   {Email$^*$:tanmay.physics@gmail.com}\\
   {Email$^{\dagger}$:muktish.physics@presiuniv.ac.in}
\end{center}
\vspace {1.0 cm}

\noindent {\bf Abstract:} A simplified model for the complex glycolytic process was historically proposed by Sel'kov. It showed the existence of stable limit cycle as an example of Poincare'-Bendixson theorem. This limit cycle is nothing but the time eliminated Lissajous plot of the concentrations of Adenosine-diphosphate (ADP) and Fructose-6-phosphate (F6P) of a normal/healthy human. Deviation from this limit cycle is equivalent to the deviation of normal physiological behaviour. It is very important to know how long a human body will take to reach the glycolytic stable limit cycle, if deviated from it. However, till now the convergence time, depending upon different initial parameter values, was not studied in detail. This may have great importance in understanding the recovery time for a diseased individual deviated from normal cycle. Here the convergence time for different initial conditions has been calculated in original Sel'kov model. It is observed that convergence time, as a function of the distance from the limit cycle, gets saturated away from the cycle. This result seems to be a physiological disorder. A possible mathematical way to incorporate this in the Selkov model, has been proposed.\\\\  

\vskip 4cm


\vskip 2cm
------------------------------------------------------------------------------------

This paper is dedicated to the memory of Prof. Dietrich Stauffer
\newpage

\noindent {\bf I. Intoduction:}
	
	Biological systems are generally very complex and hard to analyze mathematically due to presence of many variables and in most cases their exact dependence upon themselves are also unknown. However, simple models using differential equations has been used for a long time\cite{selkov,strogatz,murray,goldbeter}.
	
	Ludwig\cite{Ludwig1,Ludwig2} gave elegant picture of population biology in one dimensional dynamical system using simple differential equation. Lotka-Volterra model\cite{murray} had produced an important discovery leading to the {\it principle of competitive exclusion}. Periodic biological oscillators, by itself can be called a discipline. The rhythmic heart beat\cite{goldbeter}, circadian cycles echoing the day cycle of twenty four hours\cite{winfree1,winfree2} periodic Belousov-Zhabotinskii type oscillations in ventricle of heart\cite{winfree3} or molecular biological modeling of cycles in a cell\cite{Tyson} all having biological oscillatory motion. 
	
	In biological science, Glycolysis is an important physiological process and has been
	drawing\cite{lenzen} the attention of modern research over the last few decades.
	Glycolysis is the chain of chemical reaction through which a cell breaks glucose and assimilates it. Here the periodic behavior again plays the key role as shown both experimentally\cite{hess1,hess2} and theoretically\cite{nicolis,pye}. The role of glycolysis in the formation of endothelial tip cells during angiogenesis was studied\cite{arik} recently.  The aerobic glycolysis and its key enzimes may be used for lung cancer\cite{li}. Very recently, dynamics\cite{rendall1} of Sel'kov's oscillator and unboundedness\cite{rendall2} in solution of the model of glycolysis were thoroughly investigated.
	
	In the year 1968,  Sel'kov proposed	\cite{selkov,merkin1,merkin2} a simple two dimensional mathematical model which mimicked the periodic behavior of glycolysis process. It showed the possibility of existence of stable limit cycle in the phase space of concentration of two constituents Adenosine diphosphate (ADP) and Fructose-6-Phosphate(F6P). In a normal human body, the concentrations of ADP and F6P change in time but the time eliminated plot of ADP and F6P gives a stable closed loop called a limit cycle.  
	If a conditional trapping region is found, the Poincare'-Bendixson theorem confirms, the existence of the stable limit cycle of ADP and F6P. If the initial condition of the concentrations of ADP and F6P, are taken anywhere (other than values on the limit cycle), the dynamics itself will	bring the concentrations of ADP and F6P on the limit cycle. From the biological point of view, if by any means, the concentrations are away from the limit cycle, it will eventually settle into the limit cycle. For this reason, the time required to reach the stable limit cycle, from any arbitrary initial condition,
	is an important issue of study. This will give an idea of the time of regaining	normalcy of physiological process, if deviated.
	
	 If the initial concentrations of the constituents (ADP and F6P here) are not on the limit cycle then the amount of time required to reach the limit cycle can be named as convergence time.  
	For any point away from the limit cycle in the phase plane, concerned cell is in diseased state and convergence time represents the corresponding recovery time and has direct biological importance. The ranges of parameters in the concerned model for obtaining closed cycles are well understood. However, convergence time for different initial concentrations for the model is not well studied and is the major issue  of this  article. 
	
	It is quite natural to expect a longer recovery time for initial conditions that are further away from the limit cycle which in turn should result in longer convergence time. However, the Sel'kov model shows that convergence time as a function of distance from the limit cycle gets saturated in the regions far away from the limit cycle. This means that the degree to deceased cell do not affect the recovery time of the cell which is difficult to explain from biological perspective and limits the application of Sel'kov model to real life systems. An essential modification of the model has been proposed here, to eliminate this problem, without destroying the stable limit cycle for the model.  

The manuscript is organised as follows: after the brief introduction to the Sel'kov
model in the next section (Section-II), we have reviewed the results
(section-III) of existing
Sel'kov model and reported the results in the {\it modified} model. The paper ends with
a few words as concluding remarks in section-IV.

\vskip 1cm
	
\noindent {\bf II. Sel'kov Model of Glycolysis:}
		
		Sel'kov proposed \cite{selkov} a simplified model \cite{merkin1,merkin2} for glycolysis with only two variables as follows: \\
		
		\begin{equation}
		\begin{split}
		\frac{dx}{dt} &= -x + ay + x^2y\\
		\frac{dy}{dt} &= ~b - ay - x^2y		
		\end{split}
		\end{equation}
		
where, $x$ and $y$ stood for dimensionless concentrations of ADP and F6P, respectively and $a$ and $b$ represented two real valued positive constants. As mentioned by Sel'kov, this extremely simplified model, failed to explain more complex behaviors like  double self-oscillation seen in Glycolysis process\cite{pye}. However, it captured the essence of periodicity in the process involved and is used for the introductory mathematical model. Periodicity of the concerned model is incorporated as a stable limit-cycle appearing in the phase plane. 

It could be analytically shown that the equations incorporated a basin of attraction around a stable fixed point for some range of $a$ and $b$ values.  For a change of parameter value, the fixed point undergoes a supercritical Hopf bifurcation which changes the trajectories around fixed point from converging to diverging in nature. However, the trajectories away from the fixed point remain convergent and a limit cycle emerges in between. The region of phase space around the fixed point where all local trajectories gets trapped is called the trapping region. Poincare-Bendixson theorem
\cite{hilborn,strogatz} guaranties a periodic stable limit cycle for a range of parameter values\cite{strogatz} of $a$ and $b$ corresponding to existence of a trapping region excluding any fixed point in two dimension. Presence of limit cycle ensures the periodic feature of glycolytic process. \\\\

\vskip 1cm

\noindent {\bf III. Results:}
	
\noindent{\it {(A) Original Sel'kov Model}}
	
Although Sel'kov model leading to periodic limit cycle was well understood, comparative studies on transient states corresponding to different initial conditions do require some observations. For that the values of $a$ and $b$ are respectively chosen to be 0.08 and 0.6.Then the set of coupled differential equations are solved using 4th order Runge-Kutta method\cite{scarborough,numerical} with step-size 0.01 and a solution set $\{x_i(t),y_i(t)\}$ corresponding to initial conditions $(x_i,y_i)$ has been obtained. It is observed that solutions $\{x_i(t),y_i(t)\}$ in $t$ eliminated form eventually converges upon a closed curve called limit cycle. Points belonging to limit cycle are denoted by $\{X_k,Y_k\}$. Distance between two points $\{X_i,Y_i\}$ and $\{X_j,Y_j\}$ on the limit cycle being denoted by $d^0_{ij} = \sqrt{(X_i - X_j)^2 +(Y_i-Y_j)^2}$. Let the minimum value among $\{d^0_{ij}\}$ set be $d^0$. For a given initial condition  $(x_i,y_i)$, let us define a time dependent distance function $d_{ik}(t) = \sqrt{(x_i(t)-X_k)^2 + (y_i(t) - Y_k)^2}$ and $d_i(t) = min{\{d_{ik}(t)\}}$ for all $k$ where $min$ stands for the minimum value. If $d_i(t \geq \tau_i) \le d^0$ then $\tau_i$ stands for the convergence time onto limit cycle for initial condition  $(x_i,y_i)$.\\

For comparative study, a  $300\times300$ lattice of initial conditions $(x_i,y_i)$ ranging $0 \leq x_i \leq 3$ and $0 \leq y_i \leq 3$ have been generated. Thus lattice points are separated by a distance 0.01 both along X and Y direction. For each initial condition $(x_i,y_i)$, convergence time $\tau_i$ is being determined using above prescription. A colored image plot of $\tau_i(x_i,y_i)$ is shown in fig-1. The adjacent color bar indicates recovery time for corresponding initial condition in $x-y$ plane with warmer color showing longer and cooler color indicating shorter time range. Any initial condition being away form limit cycle symbolizes unhealthy situation for biological processes. This in turn generates expectation for longer recovery time for initial conditions away from limit cycle but as seen from fig-1, convergence time gets saturated in further regions. This contradictory observation prompted us to think for possible modification of Sel'kov model\cite{selkov,merkin1,merkin2} in glycolysis. \\\\

\noindent {\it {(B) Mathematical incorporation of physiological disorder in Sel'kov model}}  
	
	The saturation of convergence time at distant parts of $x-y$ plane can be understood from the diverging nature of the vector fields in Sel'kov model. For elimination of this problem we need to include a damping envelop in the existing model without destroying the periodic limit cycle which is the essence of Sel'kov model for glycolysis process. All these may be achieved in the following modified model:
	\begin{equation}
	\begin{split}
	\frac{dx}{dt} = f(x,y)(-x + ay + x^2y)\\
	\frac{dy}{dt} = f(x,y)(b - ay - x^2y)
	\end{split}
	\end{equation}
	where $f(x,y)$ is positive definite function with some additional conditions on partial derivatives to be discussed later.\\

	Noting that $f(x,y)>0$ everywhere and $a$, $b$ being real valued positive number this model, the same line of argument for finding the trapping region is followed that was used for the existing model\cite{strogatz} which is shown below:\\
	
	(i) Construction of trapping region:\\
	(a) Null-clines : $ \frac{dx}{dt} = f(x,y)(-x + ay + x^2y) = 0 \Rightarrow y = \frac{x}{a+x^2} $, $\frac{dy}{dt} = f(x,y)(b - ay - x^2y) = 0 \Rightarrow y = \frac{b}{a+x^2}$, since $f(x,y) \ge 0$ for over the phase-space. Since this is same as the original model, the fixed points are also unchanged at $(x^*,y^*) = (b, \frac{b}{a+b^2})$.\\
	(b) $\frac{dy}{dt} = 0$ nullcline intersects Y axis at a value $\frac{b}{a}$ which is denoted by D in Figure 2.\\
	(c) From point D, a straight line parallel to X axis is drawn upto $x=b$ which is denoted by C.\\
	(d) From C, a straight line (slope=-1) is being drawn upto nullcline $ \frac{dx}{dt} = 0$ and corresponds to line segment CB.\\
	(e) Finally, a line from B is drawn parallel to Y axis upto X axis and named as BA\\
	Thus we get a bounded region OABCD shown in Figure 2.\\
	
	(ii) Justification for Trapping region:\\
	One can observe that along different line segments, following arguments hold for direction of vector fields:\\
	(a) For OA, $\frac{dy}{dt} = f(x,y) b > 0$.\\
	(b) For OD, $ \frac{dx}{dt} = f(x,y)ay > 0 $\\
	(c) For CD, $ \frac{dy}{dt} = -f(x,y)\frac{bx^2}{a} < 0$\\
	(d) For BC, $\frac{dy}{dt} + \frac{dx}{dt} = f(x,y)(b-x) < 0 $ since, $x>b$ for this line segment and again $f(x,y) > 0$ $\Rightarrow \frac{dy}{dx} < -1$ and BC has a slope of $-1$.\\
	(e) For BA segment, $ \frac{dx}{dt} = f(x,y)(-x+ay+x^2y) <0$ since $ y < \frac{x}{a+x^2}$ along this line.\\
	
	Final point to discuss is the nature of fixed point. As we may see that assuming $f_x =\frac{\partial f}{\partial x}$, $f_y =\frac{\partial f}{\partial y}$; the Jacobin matrix corresponding to the modified model becomes:\\
	
	\begin{equation*}
	\begin{split}	
	J &= \begin{pmatrix}
	(-x+ay+x^2y)f_x+(-1+2xy)f & (-x+ay+x^2y)f_y+(a+x^2)f\\
	(b-ay-x^2y)f_x+(-2xy)f & (b-ay-x^2y)f_y+(-a-x^2)f\\
	\end{pmatrix}_{(x^*,y^*)}\\
	&= \begin{pmatrix}
	(-1+2x^*y^*)f(x^*,y^*) & (a+{x^*}^2)f(x^*,y^*)\\
	(-2x^*y^*)f(x^*,y^*) & (-a-{x^*}^2)f(x^*,y^*)\\
	\end{pmatrix}\text{(assuming $f_x,f_y$ exists at $(x^*,y^*)$)}\\
	& = f(x^*,y^*) \begin{pmatrix}
	-1+2x^*y^* & a+{x^*}^2\\
	-2x^*y^* & -a-{x^*}^2\\
	\end{pmatrix}\\
	\Rightarrow J &= f(x^*,y^*) J_0
	\end{split}
	\end{equation*}

	where $J_0$ stands for Jacobian of the original Sel'kov model (equation -1). Thus the eigenvalues of this model is $f(x^*,y^*)$ times the original model. Now, $f(x^*,y^*)$ being positive definite, we get the same sign for the eigenvalues and thus the nature of fixed points remain unchanged over the same range of parameters $(a,b)$. However the important part to notice here is that the existence of $\frac{\partial f}{\partial x}$ and $\frac{\partial f}{\partial y}$ at $(x^*,y^*)$ is required for this to be plausible. Thus it may be concluded that OABCD region excluding the neighborhood of fixed point E at $(x^*,y^*)$ is a trapping region (as stated by Poincare'-Bendixson theorem). It is schematically drawn as the colored region in Figure 3.\\

	{\it Out of infinitely many such changes, a physically satisfying model should incorporate a function that eliminates the diverging vector fields of the original model by dampening them out in regions far from limit cycle.} We have chosen the following functional form:
	\begin{equation}
	 f(x,y) =\exp\bigg[{-{\alpha}{\sqrt{(x-x^*)^2 + (y-y^*)^2}}}\bigg]
	 \end{equation}
	  
\noindent where $\alpha$ is a real valued control parameter which controls the damping. It may be noted here that $\alpha=0$ restores the original Sel'kov model (equation-1).
	 For comparison of this new model with the existing one, again convergence time $\tau_i(x_i, y_i)$ for same range initial condition$(x_i,y_i)$ has been calculated using  similar algorithm over $300 \times 300$ lattice points for $\alpha = 0.5$. Parameter values were again chosen to be $a=0.08$ and $b=0.6$. This particular choice of $a$ and $b$, gives the solution of stable limit cycle\cite{strogatz,murray}. A 
	 colored image plot of recovery time ($\tau_i$) as a function of initial ADP and F6P concentration $(x_i,y_i)$ has been shown in fig-3. Adjacent color bar here also represents the different time scales with warmer  and  cooler colors representing longer and shorter recovery time, respectively. In this color diagram it can be noticed that the recovery time does not get saturated and keeps on increasing at distant parts in the phase plane and can fit in the expectation of having larger recovery time corresponding to further disturbed initial conditions.

It may be noted here, that just by introducing a damping function $f(x,y)$ in the
Sel'kov model, one can get the realistic behaviour, i.e., distant point will take
longer time (as compared to the point closer to the limit cycle) to achieve the stable limit cycle. This does not violate any other norms of stable limit cycle but gives
realistic results which was not found in original Sel'kov model. {\it What is the reason behind it ?} In the original Sel'kov model the value of the velocity
$({{dx} \over {dt}}, {{dy} \over {dt}})$ is function of the position $(x,y)$. From
this functional form (equation-1), the magnitude of velocity is large at any point
far away from the stable limit cycle. So, the initial condition, chosen far away from
the stable limit cycle, moves faster than that chosen nearby. The functional form 
results almost the uniform value of convergence. But in the modified model
(equation-2), the
multiplicative positive function $f(x,y)$ (equation-3) could indeed slow down the flow at any
point far away from the stable limit cycle. This creates the nonuniform convergence
time $\tau_i(x_i,y_i)$ to reach the stable limit cycle. As a result, the convergence
time is larger for a point far away from the limit cycle. This is of course the aclaimed realistic result.	
	
For completeness of the discussion, line-scans of fig-1 and fig-3 are drawn for different dimensionless concentrations of F6P. Line-scans for original Sel'kov model
(equation-1) and modified Sel'kov model (equation-2) are shown in fig-4 and fig-5, respectively where the X axis shows the concentration of ADP and Y axis shows the convergence time corresponding to different choices of concentrations of F6P. In both cases, the values of F6P concentration were chosen to be 1.2, 1.3, 1.4, 1.5 and 1.6 which are plotted with different colors as shown in respective figures. Insets of both figure show the exact location across which line-scans are observed. Comparing them, Sel'kov model shows a saturation of convergence time as we move away from the limit cycle along the line-scan while the modified model shows a monotonically increasing convergence time for increment of distance from the limit cycle and serves as a better physical model.

\vskip 1cm 

\noindent {\bf IV. Concluding remarks:}

The Sel'kov model for glycolysis has been an introductory model conveying its associated periodic behavior for a long time. This celebrated model showed the existence of periodic limit cycle which arises as a consequence of Poincare'-Bendixson theorem. This limit cycle is, the Lissajous plot of the concentrations of Adenosine-diphosphate (ADP) and fructose-6-phosphate (F6P) of a normal/healthy human body. Any deviation from this limit cycle can be viewed as physiological disturbance and in such situation, it is very important to know the required recovery time. However, the convergence time, depending upon different initial parameter values, was not investigated in detail. 

 In this article, the convergence time for the different initial conditions has been calculated in original Sel'kov model. It has been observed 
surprisingly that this convergence time, studied as a function of the distance from the limit cycle, gets saturated away from the cycle. As far as knowledge of the authors is concerned, this has not been noticed before.

A possible way out has been proposed (by introducing a positive damping function
$f(x,y)$ in the original Sel'kov model) without violating any norms of Poincare'-Bendixson theorem. Our proposed modified model has shown to eliminate the problem of saturating behavior shown by recovery time.  {\it The present study is an appeal to the experimental biologists to verify such behaviours of convergence time by real data.} \\

\noindent {\bf Acknowledgements:} MA acknowledges FRPDF, Presidency University for
financial support. We thank A. D. Rendall for providing us important informations
regarding the literature of glycolysis. This paper is dedicated to the memory of
Professor Dietrich Stauffer from whom MA has learnt important numerical techniques
during (1997-98) his stay at the University of Cologne, Germany as a postdoctoral fellow.

\newpage
	
\begin{figure}[h]
		\begin{center}
			\includegraphics[height=12cm, angle=0]{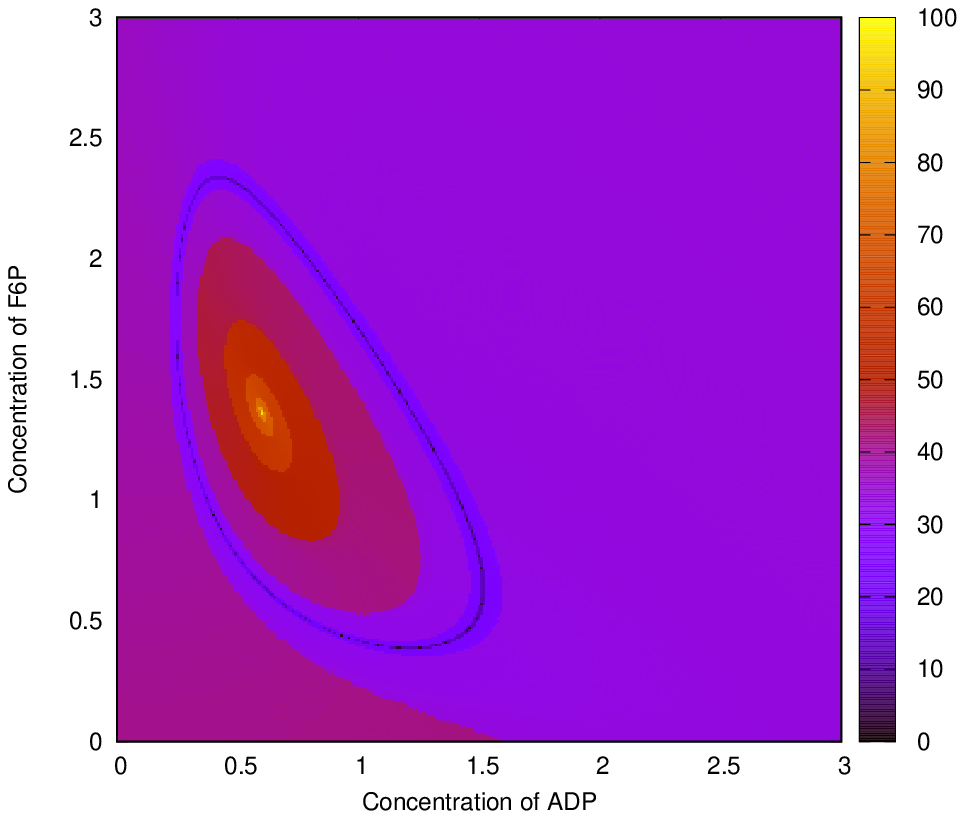}
			\caption{Image plot of convergence time $\tau_i(x_i,y_i)$ for original Sel'kov model. Color at any point is indicative of convergence time for that initial condition. Adjacent color bar is showing the exact dependence.}
			\label{Figure1}
		\end{center}
	\end{figure}

\newpage	
\begin{figure}[h]
		\begin{center}
			\includegraphics[height=12cm, angle=0]{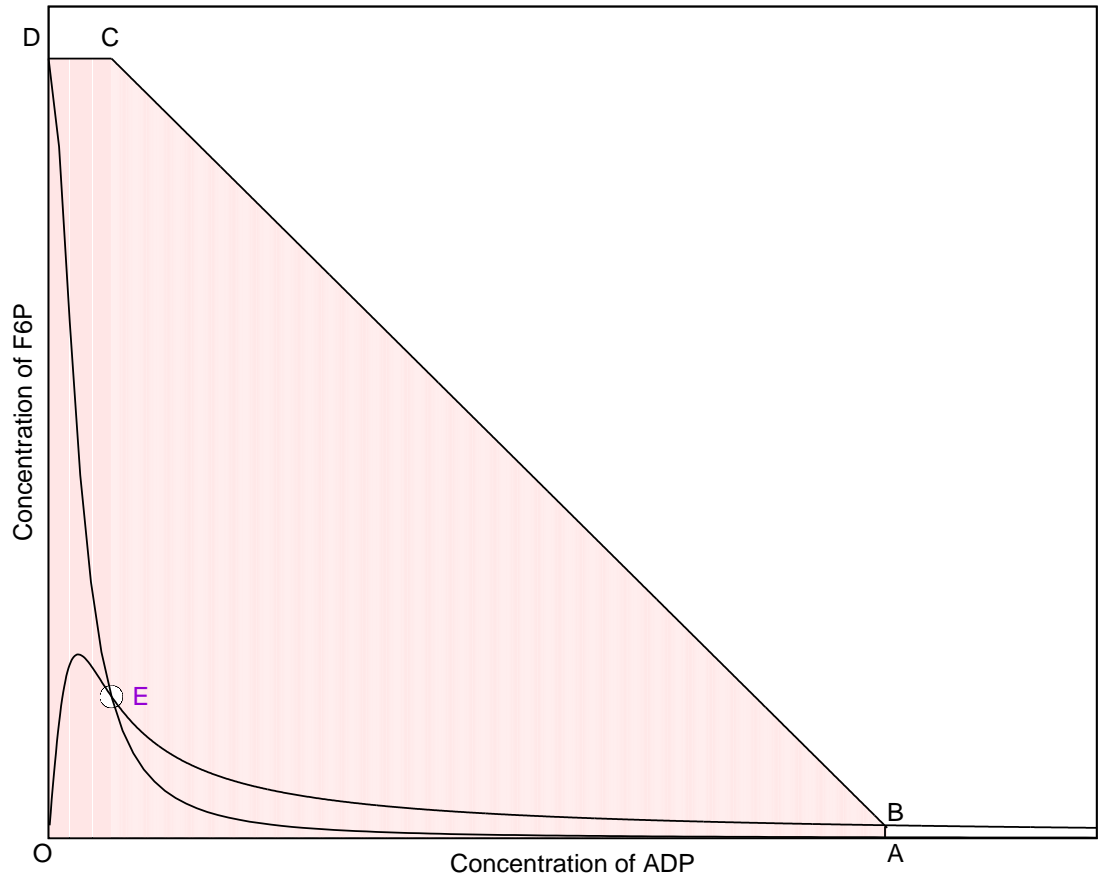}
			\caption{Trapping region for the modified Sel'kov model shown in color. It excludes the neighborhood of unstable fixed point at E.}
			\label{Figure2}
		\end{center}
	\end{figure}

\newpage	
	\begin{figure}[h]
		\begin{center}
			\includegraphics[height=12cm, angle=0]{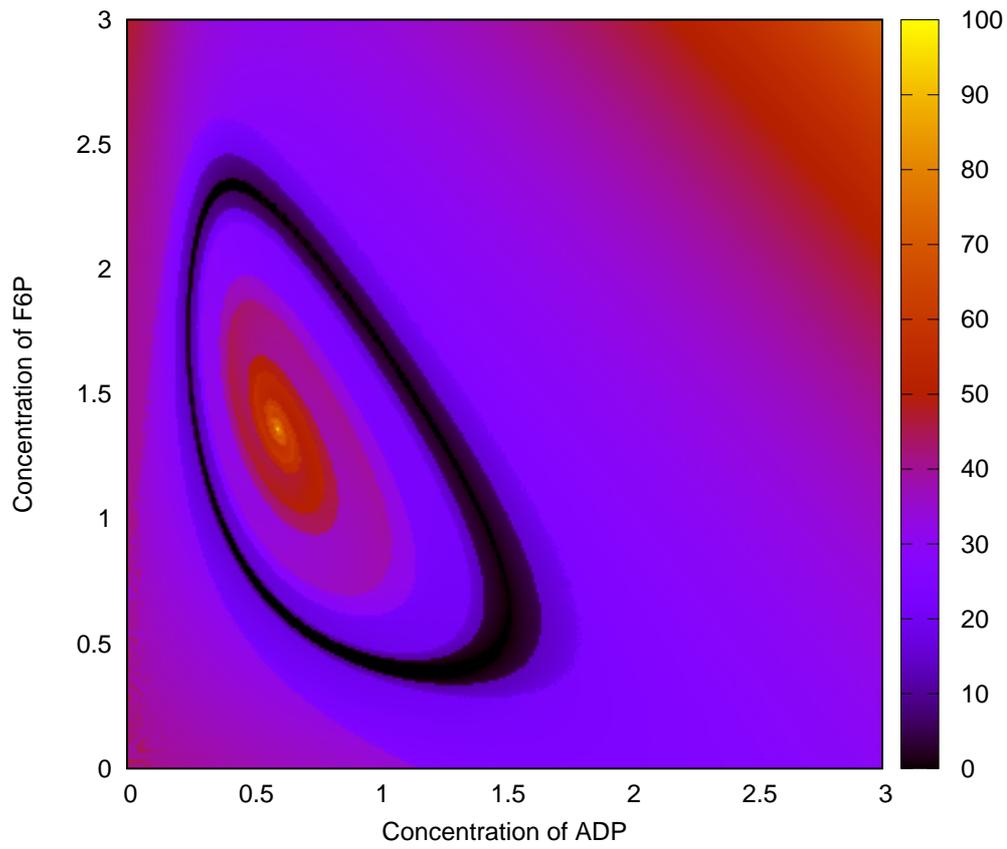}
			\caption{Image plot of the convergence time $\tau_i(x_i,y_i)$ for modified Sel'kov model with control parameter value $\alpha$ = 0.5. Color at any point is indicative of convergence time for that initial condition. Adjacent color bar is showing the exact dependence. }
			\label{Figure3}
		\end{center}
	\end{figure}

\newpage
	\begin{figure}[h]
	\begin{center}
		\includegraphics[height=12cm, angle=0]{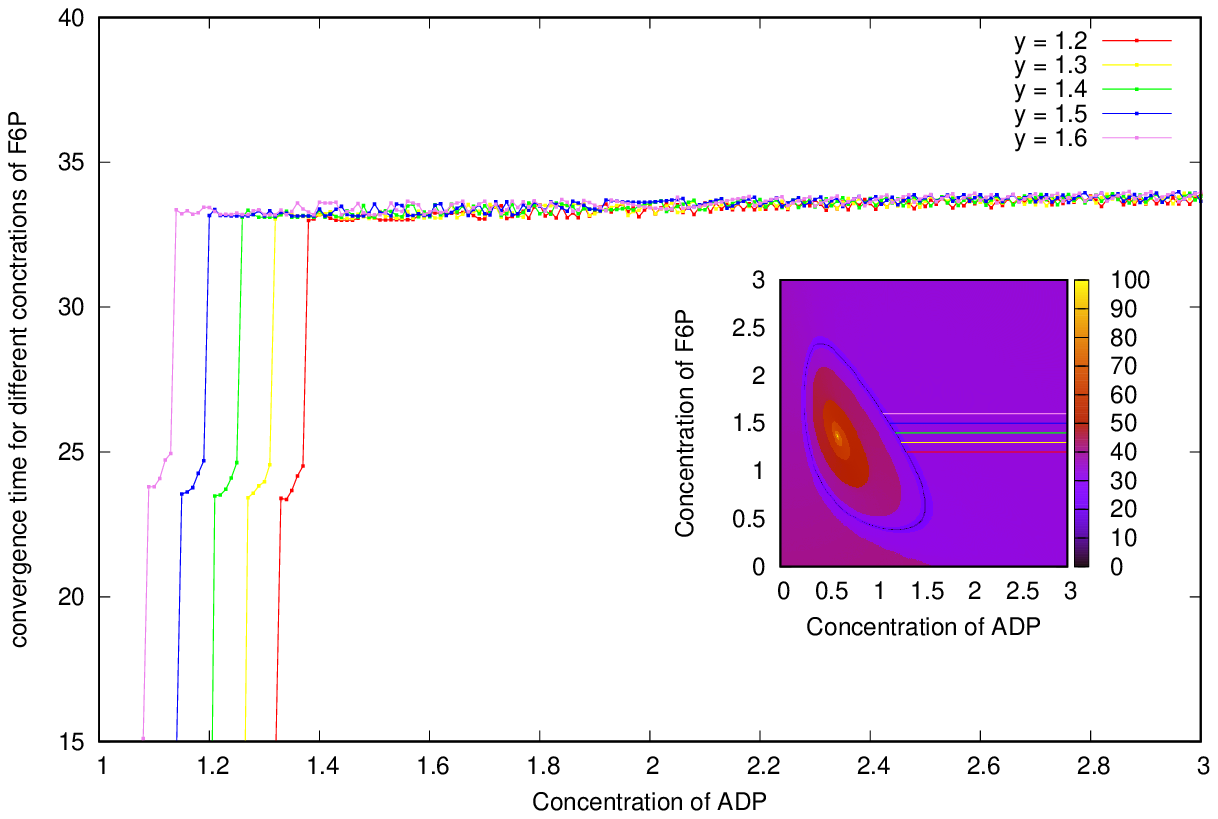}
		\caption{Line-scans of time-matrix for original Sel'kov model shown in Figure- 1 for fixed concentrations of F6P. Range of different line-scans are shown in the inset.}
		\label{Figure4}
	\end{center}
	\end{figure}

\newpage
\begin{figure}[h]
	\begin{center}
		\includegraphics[height=12cm, angle=0]{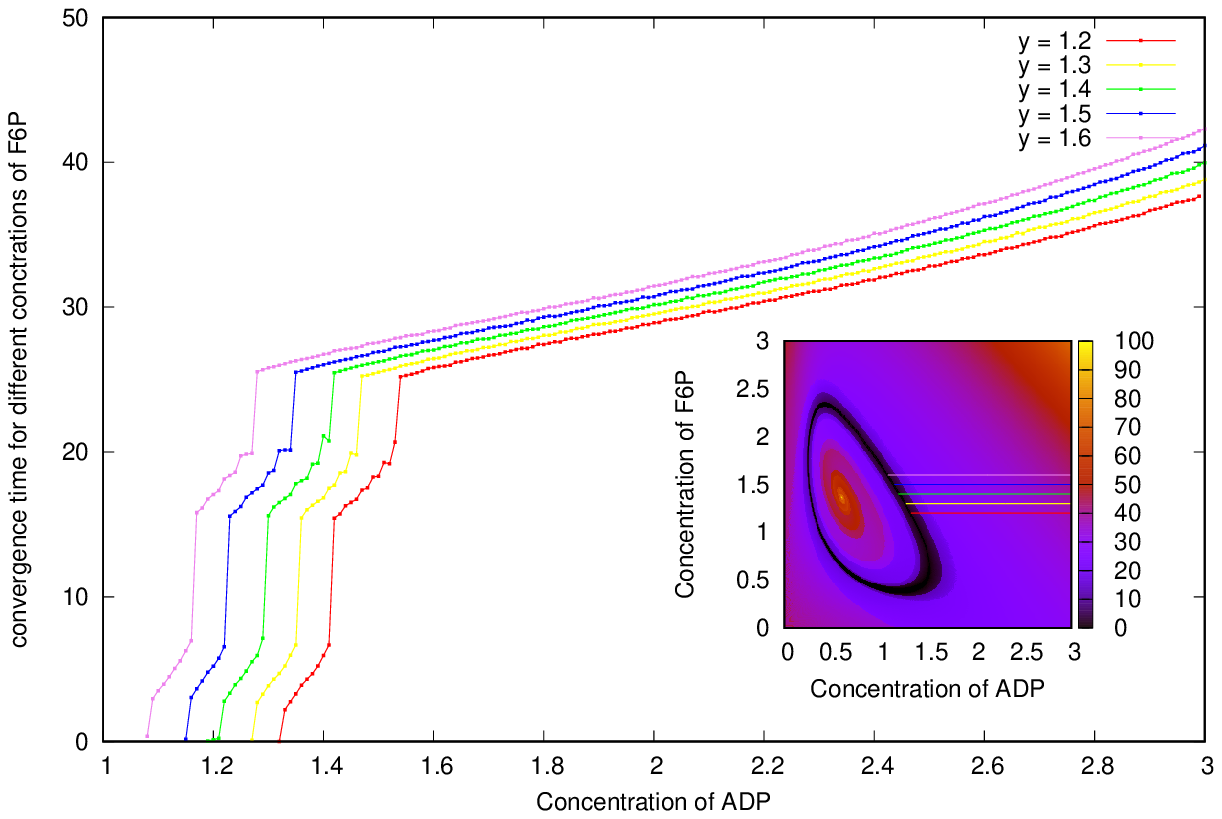}
		\caption{Line-scans of time-matrix for modified Sel'kov model shown in Figure-3 for fixed concentrations of F6P. Range of different line-scans are shown in the inset.}
		\label{Figure5}
	\end{center}
\end{figure}

\end{document}